\documentclass[twocolumn]{aastex631}
\usepackage{comment}
\usepackage{amsmath}	
\usepackage{amssymb}
\usepackage{footnote}
\usepackage{appendix}
\usepackage{upgreek}
\usepackage{color, colortbl}
\usepackage{tablefootnote}
\usepackage[graphicx]{realboxes}
\usepackage{threeparttable}
\usepackage[caption=false]{subfig}
\usepackage{gensymb}
\usepackage{xcolor,colortbl}
\usepackage{cancel}

\definecolor{Gray}{gray}{0.9}
\graphicspath{{./}{figures/}}

\newcounter{daggerfootnote}

\newcommand{\DMunits}{\,pc\,cm$^{-3}$}

\def\soft#1{\texttt{#1}}

\shorttitle{FRB\,20210117A}
\shortauthors{Bhandari et al.}

\begin{document}
\title{A non-repeating fast radio burst in a dwarf host galaxy}

\correspondingauthor{Shivani Bhandari}
\email{bhandari@astron.nl}
\author[0000-0003-3460-506X]{Shivani Bhandari}\thanks{Veni fellow}
\affil{ASTRON, Netherlands Institute for Radio Astronomy, Oude Hoogeveensedijk 4, 7991 PD
Dwingeloo, The Netherlands}
\affil{Joint institute for VLBI ERIC, 
Oude Hoogeveensedijk 4, 7991 PD Dwingeloo, The Netherlands}
\affil{CSIRO, Space and Astronomy, PO Box 76, Epping NSW 1710 Australia}
\author[0000-0002-5025-4645]{Alexa C. Gordon}
\affiliation{Center for Interdisciplinary Exploration and Research in Astrophysics (CIERA) and Department of Physics and Astronomy, Northwestern University, Evanston, IL 60208, USA}
\author[0000-0002-6895-4156]{Danica R. Scott}
\affiliation{International Centre for Radio Astronomy Research (ICRAR), Curtin University, Bentley, WA 6102, Australia}
\author[0000-0003-1483-0147]{Lachlan Marnoch}
\affiliation{School of Mathematical and Physical Sciences, Macquarie University, NSW 2109, Australia}
\affil{CSIRO, Space and Astronomy, PO Box 76, Epping NSW 1710 Australia}
\affil{Astronomy, Astrophysics and Astrophotonics Research Centre, Macquarie University, Sydney, NSW 2109, Australia}
\affil{ARC Centre of Excellence for All-Sky Astrophysics in 3 Dimensions (ASTRO 3D), Australia}
\author[0000-0002-5519-9550]{Navin Sridhar}
\affiliation{Department of Astronomy, Columbia University, New York, NY 10027, USA}
\affiliation{Theoretical High Energy Astrophysics (THEA) Group, Columbia University, New York, NY 10027, USA}
\author[0000-0003-1913-3092]{Pravir Kumar}
\affiliation{Centre for Astrophysics and Supercomputing, Swinburne University of Technology, John St, Hawthorn, VIC 3122, Australia}
\affiliation{Department of Particle Physics and Astrophysics, Weizmann Institute of Science, Rehovot 7610001, Israel}
\author[0000-0002-6437-6176]{Clancy~W.~James}
\affiliation{International Centre for Radio Astronomy Research (ICRAR), Curtin University, Bentley, WA 6102, Australia}
\author[0000-0002-9586-7904]{Hao Qiu}
\affiliation{SKA Observatory, Jodrell Bank, Lower Withington, Macclesfield, SK11 9FT, UK}

\author[0000-0003-2149-0363]{Keith W. Bannister}
\affil{CSIRO, Space and Astronomy, PO Box 76, Epping NSW 1710 Australia}
\author[0000-0001-9434-3837]{Adam T.Deller}
\affiliation{Centre for Astrophysics and Supercomputing, Swinburne University of Technology, John St, Hawthorn, VIC 3122, Australia}
\author[0000-0003-0307-9984]{Tarraneh Eftekhari}\thanks{NHFP Einstein Fellow}
\affiliation{Center for Interdisciplinary Exploration and Research in Astrophysics (CIERA) and Department of Physics and Astronomy, Northwestern University, Evanston, IL 60208, USA}
\author[0000-0002-7374-935X]{Wen-fai Fong}
\affiliation{Center for Interdisciplinary Exploration and Research in Astrophysics (CIERA) and Department of Physics and Astronomy, Northwestern University, Evanston, IL 60208, USA}
\author[0000-0002-5067-8894]{Marcin Glowacki}
\affiliation{International Centre for Radio Astronomy Research (ICRAR), Curtin University, Bentley, WA 6102, Australia}
\author[0000-0002-7738-6875]{J. Xavier Prochaska}\thanks{Simons Pivot Fellow}
\affil{University of California - Santa Cruz
1156 High St.
Santa Cruz, CA, USA 95064}
\affil{Kavli IPMU (WPI), UTIAS, The University of Tokyo, Kashiwa, Chiba 277-8583, Japan}
\affil{Division of Science, National Astronomical Observatory of Japan,2-21-1 Osawa, Mitaka, Tokyo 181-8588, Japan}
\author[0000-0003-4501-8100]{Stuart D. Ryder}
\affil{School of Mathematical and Physical Sciences, Macquarie University, NSW 2109, Australia}
\affil{Astronomy, Astrophysics and Astrophotonics Research Centre, Macquarie University, Sydney, NSW 2109, Australia}

\author[0000-0002-7285-6348]{Ryan~M.~Shannon}
\affiliation{Centre for Astrophysics and Supercomputing, Swinburne University of Technology, John St, Hawthorn, VIC 3122, Australia}
\author[0000-0003-3801-1496]{Sunil Simha}
\affil{University of California - Santa Cruz
1156 High St.
Santa Cruz, CA, USA 95064}

\begin{abstract}
We present the discovery of as-of-yet non-repeating Fast Radio Burst (FRB), FRB\,20210117A, with the Australian Square Kilometer Array Pathfinder (ASKAP) as a part of the Commensal Real-time ASKAP Fast Transients (CRAFT) Survey. 
The sub-arcsecond localization of the burst led to the identification of its host galaxy at a $z=0.214(1)$. This redshift is much lower than what would be expected for a source dispersion measure (DM) of 729\,pc\,cm$^{-3}$, given typical contributions from the intergalactic medium and the host galaxy.
Optical observations reveal the host to be a dwarf galaxy with little on-going star formation, very different to the dwarf host galaxies of known repeating FRBs 20121102A, and 20190520B. We find an excess DM contribution from the host and attribute it to the FRB's local environment. We do not find any radio emission from the FRB site or host galaxy. The low magnetized environment and lack of a persistent radio source (PRS) indicate that the FRB source is older than those found in other dwarf host galaxies, and establish the diversity of FRB sources in dwarf galaxy environments. We find our observations to be fully consistent with the hypernebula model, where the FRB is powered by accretion-jet from a hyper-accreting black hole. Finally, our high-time resolution analysis reveals burst characteristics similar to those seen in repeating FRBs. We encourage follow-up observations of FRB\,20210117A to establish any repeating nature. 
\end{abstract}


\section{Introduction} \label{intro}
Fast radio bursts (FRBs) are nano- to milli-second duration pulses of coherent radio emission with dispersion measures (DM) exceeding the maximum expected contribution from the Milky Way along a given line of sight \citep{frbcat}. The majority of the published sample of $>600$ FRBs are dominated by non-repeating events; only 4\% of FRB sources are observed to emit repeating bursts \citep{chimecat}. While the fundamental relationship between repeating and non-repeating FRBs is unknown, the growing sample reveals statistical differences in the burst properties of the two speculative populations \citep{Pleunis+21}. There are, however, no significant differences between the galaxies hosting repeating and non-repeating FRBs \citep{Bhandari+22}. The localized sample of 22 FRBs mostly comes from the outskirts of their host galaxies at redshifts ranging from less than 0.001 to 1.016 and have diverse host and local environments \citep{Tendulkar17,Ravi+19,Marcote+20,Bhandari_2020,Heintz2020,Fong+21,Niu+22,Bhandari+22,Kirsten+22,Ryder+22,Ravi+22}.

The first repeating FRB\,20121102A \citep{2016Natur.531..202S} is localized to a low-metallicity dwarf host galaxy with a high specific star formation rate at $z=0.192$ \citep{Tendulkar17}. The burst was found to be co-located with a compact persistent radio source (PRS; $<$0.7\,pc in size) suggesting that the FRB source is embedded in a  radio nebula \citep{VLAlocalisation,Marcote17}. Also, the repeat bursts were observed to have exceptionally high ($\sim 10^{5}$\,rad\,m$^{-2}$) and highly variable rotation measure \citep{MichilliRM,Hilmarsson+21}. The properties of the local environment and host galaxy of FRB\,20121102A led to a concordant model for FRBs in which bursts are produced by young magnetars, themselves produced in superluminous supernovae or long gamma-ray bursts \citep{Margalit+18}. Alternatively, the PRS can also be self-consistently explained by an accreting compact object engine \citep{Sridhar&Metzger22, 2022arXiv220100999C}. 

More recently, the repeating FRB\,20190520B was discovered using the FAST radio telescope. The observed DM of 1202\,pc\,cm$^{-3}$ would imply a redshift of $z \gtrsim 1$ \citep{JP+20}. 
Surprisingly, however, the localization of
the FRB and optical observations revealed a dwarf host galaxy at $z = 0.241$, making this source the FRB with the highest host DM contribution of DM$_{\rm host} = 903_{-111}^{+72}$\,pc\,cm$^{-3}$ \citep{Niu+22}. This is
unlikely to be due to the interstellar medium of the host galaxy, but rather more plausibly from the local environment of the source. This host DM is a factor of $\sim 5$ larger than what is observed
for FRB host galaxies \citep{James+22b} and a factor of a few beyond what is estimated for FRB\,20121102A \citep{Tendulkar17}.
Interestingly, similar to FRB\,20121102A, FRB\,20190520B is co-located with a PRS (the second only ever to be found). Furthermore, these two FRBs are among the active repeating sources and are also linked with PRSs, implying that these may be the characteristics of young and active FRB sources surrounded by dense and magnetized plasma. 

Alternatively, other FRBs have been found in massive and moderately star-forming galaxies lacking a strong magnetic environment and radio nebula. It is possible that such sources are relatively older or live in less dense environments leading to an underluminous PRS \citep{Margalit+19, Sridhar&Metzger22}. Also, a CHIME/FRB repeating source FRB\,20200120E was recently localized to a globular cluster in the galaxy M81, revealing a very different local environment for this source \citep{Kirsten+22}.

It is appealing to explain the wide variety of FRB environments using a connected mechanism, which is typically attributed to either the source age or the source formation channel.
In either case, a knowledge of the environment surrounding a larger sample of FRBs is the key to understanding this potential connection. The presence or absence of a PRS or radio emission from star formation, and how it correlates with FRB properties such as repetition rate, dispersion measure due to the host galaxy, rotation measures etc, is thus critical. 

We present the discovery of the apparently non-repeating FRB\,20210117A with the Australian Square Kilometre Array Pathfinder (ASKAP) and its localization to a dwarf galaxy in this paper. Section 2 describes the discovery as well as the properties of the host galaxy. Section 3 presents the high time resolution analysis of the burst. Section 4 describes radio follow-up observations made to look for a PRS and repeating bursts from the source of FRB\,20210117A. Section 5 discusses the implications of our findings, and Section 6 provides a summary. 

\section{Discovery of FRB\,20210117A}
The burst was detected on 2021 January 17 UT 07:51:21.277 in the real-time CRAFT incoherent sum search observations using 26 ASKAP dishes at a centre frequency of 1271.5\,MHz spanning a bandwidth of 336\,MHz. These observations were carried out simultaneously with the Rapid ASKAP Continuum Survey \citep[RACS]{RACS} observation. The burst, however, was not detected in a 10-s commensal ASKAP snapshot taken during a 15-minute RACS pointing. In CRAFT data, the burst had a maximum S/N of 27.1 in the outer ASKAP beam 02 and was also detected in beams 01 and 07, with S/N of 16.3 and 4.7 respectively, of the closepack36 beam footprint pattern (see  \citealp{Shannon2018}). The burst has a fluence of 
$36^{+28}_{-9}$\,Jy\,ms and a structure-maximized DM of $729.1_{-0.23}^{+0.36}$\,pc~cm$^{-3}$, which is derived using the method described in \citet{Sutinjo23}. 
The k-corrected isotropic-equivalent spectral energy of the burst is derived using:
\begin{equation}
   E_\nu = \frac{4\pi\,D_L(z)^{2}}{(1+z)^{2+\alpha}}\,F_{\nu}, 
\end{equation}
where $D_L(z)$ is luminosity distance, $F_{\nu}$ is burst fluence and $\alpha$ is the spectral index ($F \propto \nu^{\alpha}$) \citep{James+22}. We use a default value of $\alpha = -1.5$ \citep{Macquart2019} and derive $ E_\nu = 4.6\times 10^{31}$\,erg\,Hz$^{-1}$. We note that any beaming of the FRB can reduce this energy budget by a factor of $\Delta \Omega/4\pi$, where $\Delta \Omega$ is the unknown beaming solid angle. Furthermore, if beaming is invoked to reduce the energy of a burst, it implies that such bursts are more numerous, as we only see a fraction of them. Other measured and derived properties of FRB\,20210117A are listed in Table\,\ref{tab:measurements}.

The detection in the real-time system triggered a download of 3.1\,s of voltages around the time of the FRB. Using the standard CRAFT post-processing pipeline \citep{Day2020}, we imaged both the FRB and the continuum sources visible in the field. The FRB was detected with a significance of 50\,$\sigma$, leading to a statistical positional precision of $\sim 0.1''$ in R.A. and Decl. We used the method described in \citet{Day+21} to estimate the systematic uncertainties by
identifying 7 compact sources greater than 7$\sigma$ in the 3-sec field image and comparing them to their counterparts in the RACS radio image. We obtain an offset correction of $0.02''\pm0.08''$ in R.A. and $0.01''\pm0.08''$ in Decl. 
The final burst position is R.A.(J2000): 22h39m55.015s and Decl.(J2000):\,$-$16$^{\circ}$09$'$05.45$''$ with an uncertainty of $0.13''\times0.12''$. 
\begin{figure*}
\begin{center}
\begin{tabular}{ll}
\includegraphics[width=0.34\textwidth]{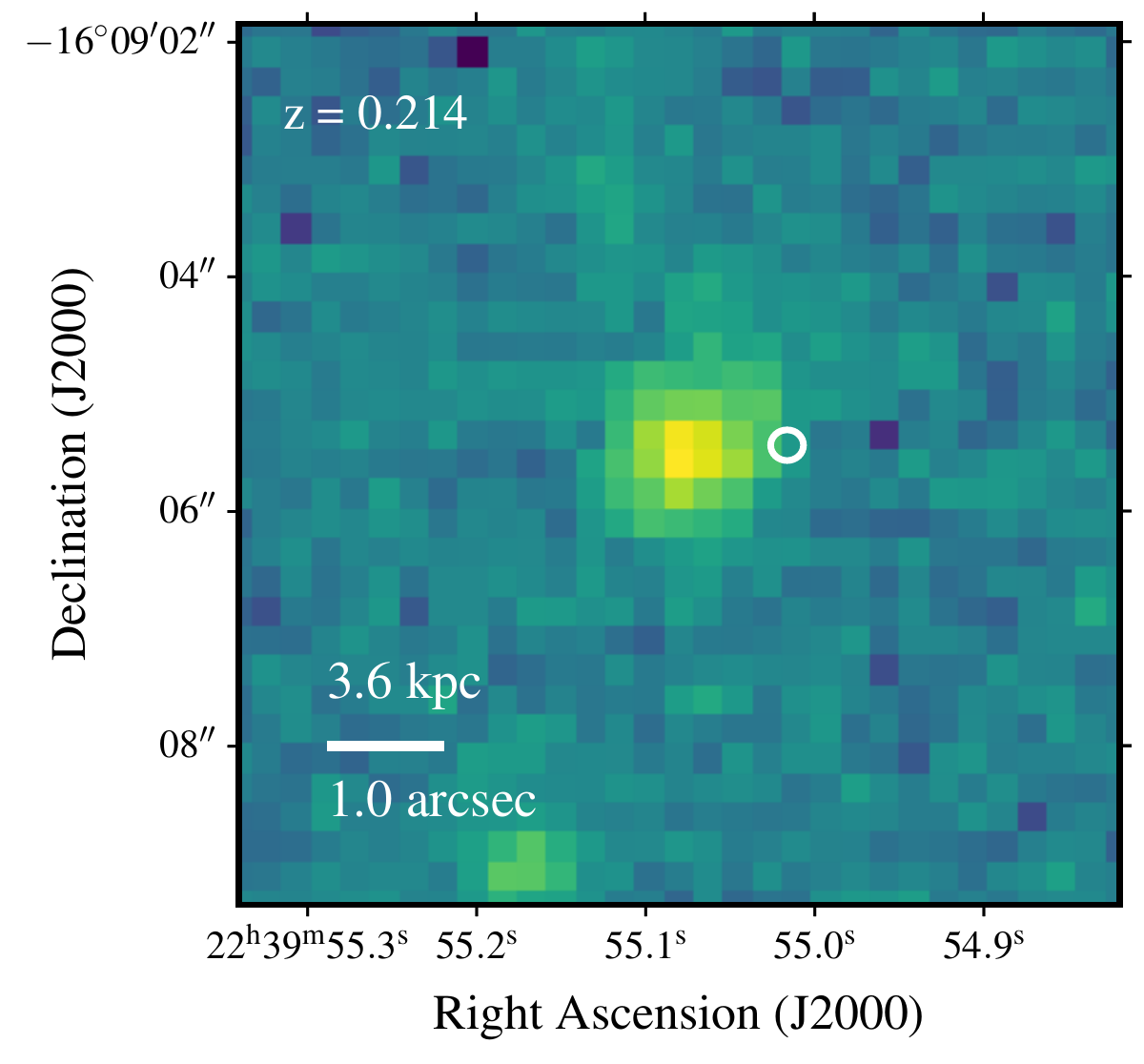} &
\includegraphics[scale=0.4]{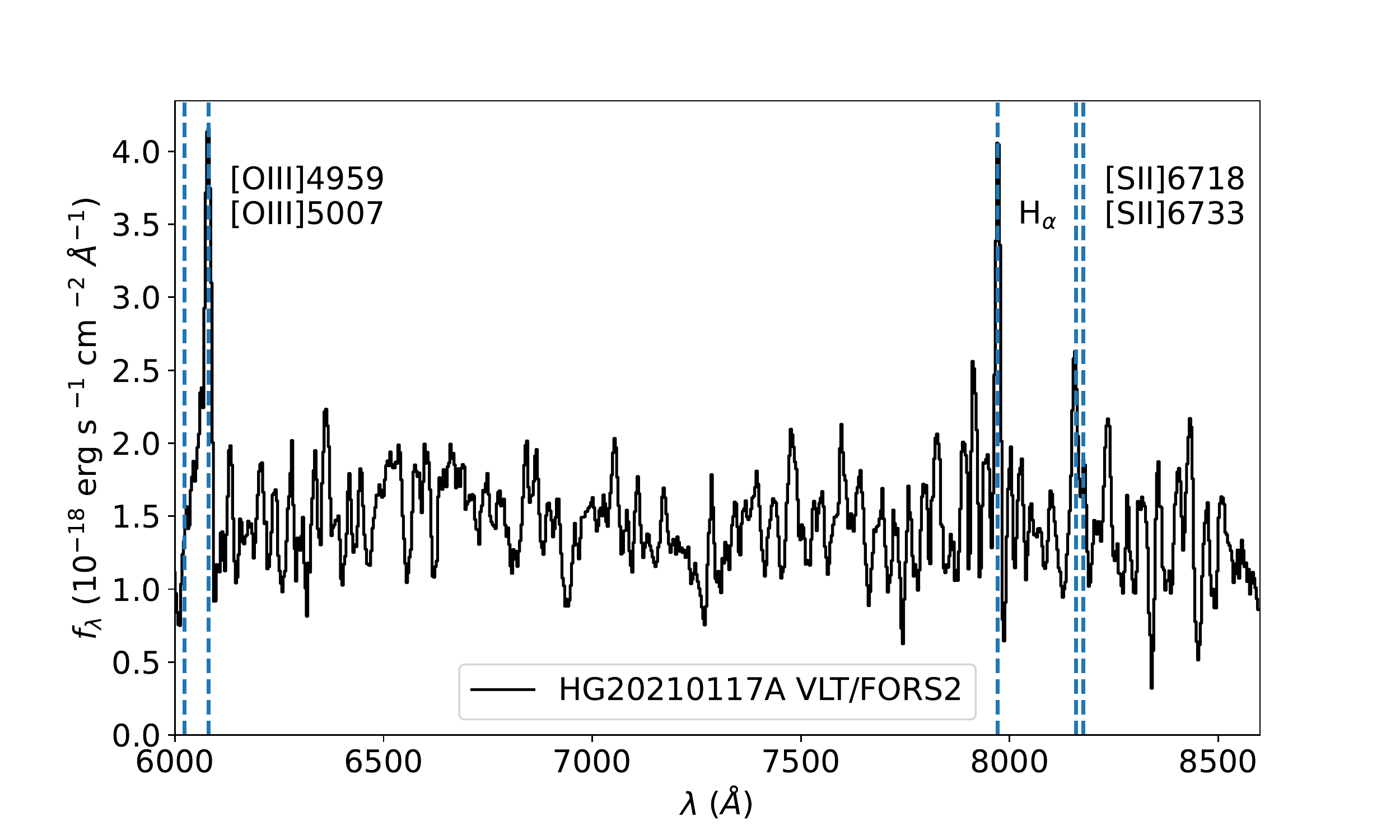} 
\end{tabular}
\caption{Left: $I$-band VLT/FORS2 image of the host galaxy of FRB\,20210117A overplotted with the position of the burst. The white circle represents the total uncertainty ($1\sigma$) in the FRB position. 
Right: VLT/FORS2 spectrum of the FRB\,20210117A host galaxy which is used to estimate the redshift of the host to be $z=0.214$. }
\label{fig:host}
\end{center}
\end{figure*}

\subsection{Host galaxy of FRB\,20210117A}
On 2021 June 10/11, we used the Keck/DEIMOS to image the field in $r$-band. The data revealed a faint galaxy with $r\sim23$, 
coincident with the position of the burst. 
We performed a Probabilistic
Association of Transient Hosts
\citep[PATH;][]{path} analysis
which yielded a $P(O|x) = 0.9984$ posterior probability that this source is the host of FRB\,20210117A. 

On 2021~June~12~UT, additional imaging observations in the $g$- and $I$-bands were obtained with the FORS2 instrument mounted on Unit Telescope 1 (UT1) of the European Southern Observatory's Very Large Telescope (ESO VLT). The images were processed as described by \citet{Marnoch2020}: debiasing and flatfielding was carried out using ESOReflex\footnote{\url{https://www.eso.org/sci/software/esoreflex/}} \citep{ESOReflex}; mosaicing with Montage\footnote{\url{http://montage.ipac.caltech.edu/}} \citep{Montage}; and astrometric calibration using a local installation of Astrometry.net\footnote{\url{http://astrometry.net/}} \citep{Astrometry} incorporating the Gaia \citep{GaiaDR2} catalog; this results in a precision (calculated as the the RMS of the offsets of imaged stars from counterparts in Gaia DR3) of $\sim0.07''$ for both bands. The $g$-band image was calibrated photometrically against DR2 of the DELVE catalogue \citep{DELVE}, and the $I$-band using the FORS2 Quality Control archive. The total integration times and image quality were 5000/900~sec and 0.70/0.65~arcsec in $g$/$I$, respectively. 
Further imaging was acquired on 2022~June~10~UT with the HAWK-I instrument, on UT4 of the ESO VLT, in $J$, $H$ and $Ks$ bands. ESOReflex was used for the debiasing, flatfielding and coaddition of the images, while photometric calibration was performed against the 2MASS Point-Source Catalog \citep{2MASS}. The astrometric calibration was performed using the same procedure as FORS2. Each band was observed for a total integration time of 750 seconds.
See Table\,\ref{tab:measurements} for photometric details.


\subsubsection{Host galaxy spectrum}
Having identified the most likely host galaxy in these images (Fig.~\ref{fig:host}), follow-up spectroscopy using FORS2 with a $1''$ slit, the GRIS300I grism and OG590 order sorting filter was obtained on 2021~Sep~6~UT. This yielded wavelength coverage of 600--1100\,nm at a resolution of 660. The total on-source exposure time was 2600~sec.

The spectrum was reduced with the Python Spectroscopic Data Reduction Pipeline \citep[\texttt{PypeIt};][]{Prochaska2020}. \texttt{PypeIt} performed flat-fielding, bias subtraction, wavelength calibration, and spectral extraction using the standard default parameters. The spectrum was then flux calibrated using the spectrophotometric standard star EG21 which was observed on 2021 Sep 2 UT. The two 1300\,s exposures were combined via 1D coaddition and scaled to match the Keck/DEIMOS $r$-band flux. Finally, the spectrum was telluric-corrected using the Paranal VIS 4900 atmospheric grid and corrected for extinction using the \citet{Calzetti2001} extinction law. A detection of H$\alpha$, [{S\,{\sc ii}}] doublet and [{O\,{\sc iii}}] doublet spectral lines confirmed the redshift of the host to be $z=0.214(1)$. No other spectral lines are apparent in the data.
\subsubsection{Stellar population modeling }
To determine the stellar population properties of the host galaxy, the stellar population synthesis modeling code \texttt{Prospector} \citep{Johnson2021} was used. The observed photometry and spectroscopy were jointly fit using the stellar population synthesis library \texttt{python-fsps} \citep{Conroy2009, Conroy2010}. We assume a Kroupa initial mass function (IMF) \citet{Kroupa01} and \citet{KriekandConroy13} dust attenuation curve. Additional assumed priors include a ratio on dust attenuation between old and young stars, mass-metallicity relationship \citep{Gallazzi2005}, and a continuity non-parametric star formation history (SFH, \citet{Leja2019}) using 8 age bins.  Several spectroscopic calibration parameters were used including a spectral smoothing parameter, a parameter to normalize the spectrum to the photometry, a pixel outlier model to marginalize over poorly modeled noise, and a jitter model to inflate the noise in all spectroscopic pixels to ensure a better fit between the model and observed spectrum. A 12$^{\rm th}$ order Chebyshev polynomial was then used to fit the spectral continuum. Our assumed model, as described above, was then sampled using the dynamic nested sampling routine \texttt{dynesty} \citep{Speagle2020} to produce the posterior distributions of stellar population parameters.

The resulting model reveals a dwarf galaxy with a stellar mass of log(M$_*$/M$_{\odot}) = 8.56^{+0.06}_{-0.08}$ and mass-weighted age of $5.06^{+0.91}_{-1.34}$ Gyr (Gordon et al 2023., in prep.), a metric less biased by the youngest and brightest stars in the galaxy compared to traditional light-weighted ages \citep{Conroy2013}. The host has a low current star formation rate (SFR) with an average SFR over the past 100\,Myr of $0.014^{+0.008}_{-0.004}$ M$_{\odot}$ yr$^{-1}$. These values and other host properties are reported in Table \ref{tab:measurements}. {We note that as \texttt{prospector} is a Bayesian inference code, the uncertainties on the stellar parameters correspond to the 68\% confidence intervals on the posteriors, given all of the priors for the assumed model.
\startlongtable
\begin{deluxetable*}{lc}
\tablewidth{0pc}
\tablecolumns{3}
\tablecaption{Measured and derived properties FRB\,20210117A and its host galaxy. \label{tab:measurements}}
\tabletypesize{\footnotesize}
\startdata 
    \textbf{Measured burst properties} \\
         Arrival time at 1271.5 MHz  & 2021-01-17-07:51:21.277 \\
         S/N  & 27.0 \\
         Structure-maximized DM (pc\,cm$^{-3}$) & $ 729.1_{-0.23}^{+0.36}$\\
         DM$_{\rm ISM}~\rm NE2001 $ (pc\,cm$^{-3}$) & 34 \\
         DM$_{\rm ISM}~\rm YMW16 $ (pc\,cm$^{-3}$) & 23 \\
         DM$_{\rm cosmic}$(pc\,cm$^{-3}$) & $\sim184$  \\
         DM$_{\rm host}$(pc\,cm$^{-3}$) & $\sim460$  \\
         RA (J2000) & 22h39m55.015(9)s  \\ 
         Dec (J2000) & $-$16$^{\circ}$09$'$05.45(12)$''$ \\  
         Fluence (Jy\,ms) & $36^{+28}_{-9}$  \\
         Peak1 pulse width (ms)\tablenotemark{$*$} & $0.14 \pm 0.01$ \\
         Peak2 pulse width (ms)\tablenotemark{$*$} & $0.17 \pm 0.02$ \\
         Precursor component pulse width (ms)\tablenotemark{$*$} & $0.53 \pm 0.03$  \\
         Scattering time ($\tau_{1.2\,\rm GHz}$) (ms) & $0.33 \pm0.02$ \\
         Rotation measure (RM) (rad~m$^{-2}$) & $43 \pm0.6$ \\
         Spectral energy density (erg\,Hz$^{-1}$)   & 4.6$\times$10$^{31}$ \\
         PRS luminosity (L$_{6\,\rm GHz}$) (W\,Hz$^{-1}$) & $<1.5\times 10^{21}$   \\
         \\
         \textbf{Host galaxy properties} \\
         RA (J2000) & 22h39m55.07(2)s  \\ 
         Dec (J2000) & $-$16$^{\circ}$09$'$05.37(2)$''$ \\
        Redshift & 0.214(1)  \\
        $g$ (AB\,mag) & $23.60 \pm 0.02$ \\
        $r$ (AB\,mag) & $22.97 \pm 0.04$ \\
        $I$ (AB\,mag) &  $22.23 \pm 0.05$\\
        $J$ (AB\,mag) & $22.69 \pm 0.08$ \\
        $H$ (AB\,mag) & $22.94 \pm 0.1$ \\
        $K$ (AB\,mag) &  $22.80 \pm 0.1$\\
        $u-r$(restframe) & $1.0 \pm 0.1$ \\
        M$_{\rm r}$ (restframe) & $-17.23 \pm 0.05$ \\
        log(M$_*$/M$_{\odot}$)
        & $8.56^{+0.06}_{-0.08}$ \\
        100 Myr SFR (M$_{\odot}$ yr$^{-1}$) & $0.014^{+0.008}_{-0.004}$ \\
        log(sSFR) (yr$^{-1}$) & $-10.4$\\
        Mass-weighted age (Gyr)  
        & $5.06^{+0.91}_{-1.34}$ \\
        Projected offset from galaxy center (kpc) & $2.8\pm 0.4$\\
 \tableline
\enddata 
\tablenotetext{*}{Reported widths are $1\sigma$ of the Gaussian}
\end{deluxetable*}

\subsection{Excess host DM}
The observed DM of the FRB can be divided into contributions from various components as
\begin{equation}
\begin{split}
    \rm DM_{obs} = DM_{MW,ISM} + DM_{MW,halo} + 
    DM_{EG}
    \\
    \rm DM_{EG} = DM_{cosmic} + \frac{DM_{host}}{1+z}. \label{eq:dm}
\end{split}
\end{equation}
Here DM$_{\rm MW,ISM}$ and DM$_{\rm MW,halo}$ are the contributions due to the Milky Way's interstellar medium and halo. These are estimated to be $34$\,pc\,cm$^{-3}$ and $23$\,pc\,cm$^{-3}$ from the Galactic models of NE2001 \citep{ne2001} and YMW16 \citep{YMW16}, respectively and DM$_{\rm MW,halo}$ is assumed to be $50$\,pc\,cm$^{-3}$ \citep{2019MNRAS.485..648P}. DM$_{\rm EG}$ refers to the extragalactic DM which is compose of the contributions due to the IGM/foreground halos along the FRB sightline (DM$_{\rm cosmic}$) and the host galaxy of FRB (DM$_{\rm host}$). DM$_{\rm cosmic}$ is estimated to be $183$\,pc\,cm$^{-3}$ using the Macquart (DM-z) relation \citep{JP+20}. After subtracting the respective contributions from the Milky Way (using NE2001) and IGM from the observed DM of the FRB, we find DM$_{\rm host}$ to be $\sim 460$\,pc\,cm$^{-3}$, which is greater than what has been observed for ASKAP-localized FRBs (a median of DM$_{\rm host} = 186^{+59}_{-48}$\,pc\,cm$^{-3}$ \citep{James+22b}). A much lower value is possible if the sightline exhibits
a higher DM$_{\rm cosmic}$ value than typical; 
see Simha et al. (in prep) for such a test hypothesis. When we include the variation in DM$_{\rm cosmic}$ from \citet{JP+20} with a feedback parameter $F=0.32$ into Eq.~\ref{eq:dm}, we produce a distribution for DM$_{\rm host}$. Scaling this to the host galaxy rest frame by $1+z$ as per \citet{Ryder+22} produces Figure~\ref{fig:pdmhost}, where the rest-frame DM is compared to other FRBs with large DM$_{\rm host}$ contributions. Using this method, we estimate the median rest-frame DM$_{\rm host}$ for FRB\,20210117 to be $595^{+55}_{-24}$\,pc\,cm$^{-3}$.

\citet{James+22} demonstrated that it is critical to consider observational biases in a survey because they can result in an inversion of the Macquart relationship after a certain DM value. Using their P(DM$_{\rm EG}$,z) grid for the CRAFT/ICS survey, we calculate the probability distribution function (pdf), P(DM$_{\rm EG}\mid $z), given the redshift of FRB\,20210117A. The pdf is presented in Fig.\,\ref{fig:dmz_host}, which reaches its maximum at an extragalatic DM of $182$\,pc\,cm$^{-3}$, with $1\sigma$ confidence interval spanning $176-496$\,pc\,cm$^{-3}$. We also show the pdf of the DM due to the IGM and the extragalactic DM, both of which are free of any instrumental biases. 

The host DM contribution can be probed by optical studies. 
We use the H$\alpha$ flux measurement from the spectrum of the host to constrain host DM. We measure $F_{\rm H\alpha} = 1.7 \times 10^{-17}\,\rm erg\,cm^{-2}\,s^{-1}$ and use it to derive the H$\alpha$ luminosity of L$_{\rm H\alpha}=2.3 \times 10^{39}$\,erg\,s$^{-1}$. 
Dwarfs of the Magellanic type range in size from 1 to 5 kpc \citep{Kaisin+12}. For simplicity, we assume the size of the dwarf host galaxy to be 3\,kpc as we are unable to fit a sersic profile due to galaxy's unresolved nature. 
Thirdly, H$\alpha$ luminosity is proportional to $\int n_{\rm e}^{2}~{\rm d}V$ because it is a tracer of ionized hydrogen, implying that the free electron density is proportional to the square root of the total H$\alpha$ luminosity emitted by the host galaxy, $n_{\rm e} \propto \sqrt{L_{\rm H\alpha}/V}$ \citep{Xu2015}, where $V$ is the volume of a sphere. We note that this assumes the volume of galaxies like the MW and dwarfs are uniformly ionized. 
According to statistics of Milky Way-type galaxies \citep{2004A&A...414...23J}, the total $L_{\rm H\alpha}$ from the Milky Way is $\sim10^{40}$\,erg\,s$^{-1}$, and the size of the MW is 30\,kpc. Finally, using the above relation, we obtain,
\begin{equation}
\begin{split}
\begin{aligned}
&\frac{n_{\rm e, host}}{n_{\rm e, MW}} \propto \sqrt{\frac{L_{\rm H \alpha, host}/V_{\rm host}}{L_{\rm H\alpha, MW}/V_{\rm MW}}}, \\
&\frac{{\rm DM}_{\rm host}}{{\rm DM}_{\rm MW}} = \frac{n_{\rm e, host}}{n_{\rm e, MW}}. \frac{l_{\rm host}}{l_{\rm MW}}, \\
&{\rm DM}_{\rm host} \propto {\rm DM}_{\rm MW} \left(\frac{L_{\rm H\alpha, host}}{L_{\rm H\alpha, MW}}\right)^{1/2}  \left(\frac{R_{\rm MW}}{R_{\rm host}}\right)^{3/2} \left(\frac{l_{\rm host}}{l_{\rm MW}}\right). 
\end{aligned}
\end{split}
\end{equation}
Here $l_{\rm host}$ and $l_{\rm MW}$ are the path lengths along the host and the Milky Way, which are assumed to be twice the effective radius of the galaxy.
Using the DM$_{\rm MW}$ contribution from NE2001, we estimate DM$_{\rm host}$ to be $\sim$60\,pc\,cm$^{-3}$. We are unable to estimate the host's inclination angle because the galaxy is barely resolved in our observations. Nevertheless, we note that in a simulation to model the DM due to dwarf galaxies, \citet{Xu2015} find the DM to be $11-12$\,pc\,cm$^{-3}$, $22-24$\,pc\,cm$^{-3}$, and about $100$\,pc\,cm$^{-3}$ for inclination angles of 0, 60, and 90 degrees respectively. Thus, we conclude that the  
host ISM alone cannot dominate the excess DM observed along the FRB sightline, and that the excess DM must come from the local environment of FRB\,20210117A.

\begin{figure}
\includegraphics[width=0.5\textwidth]{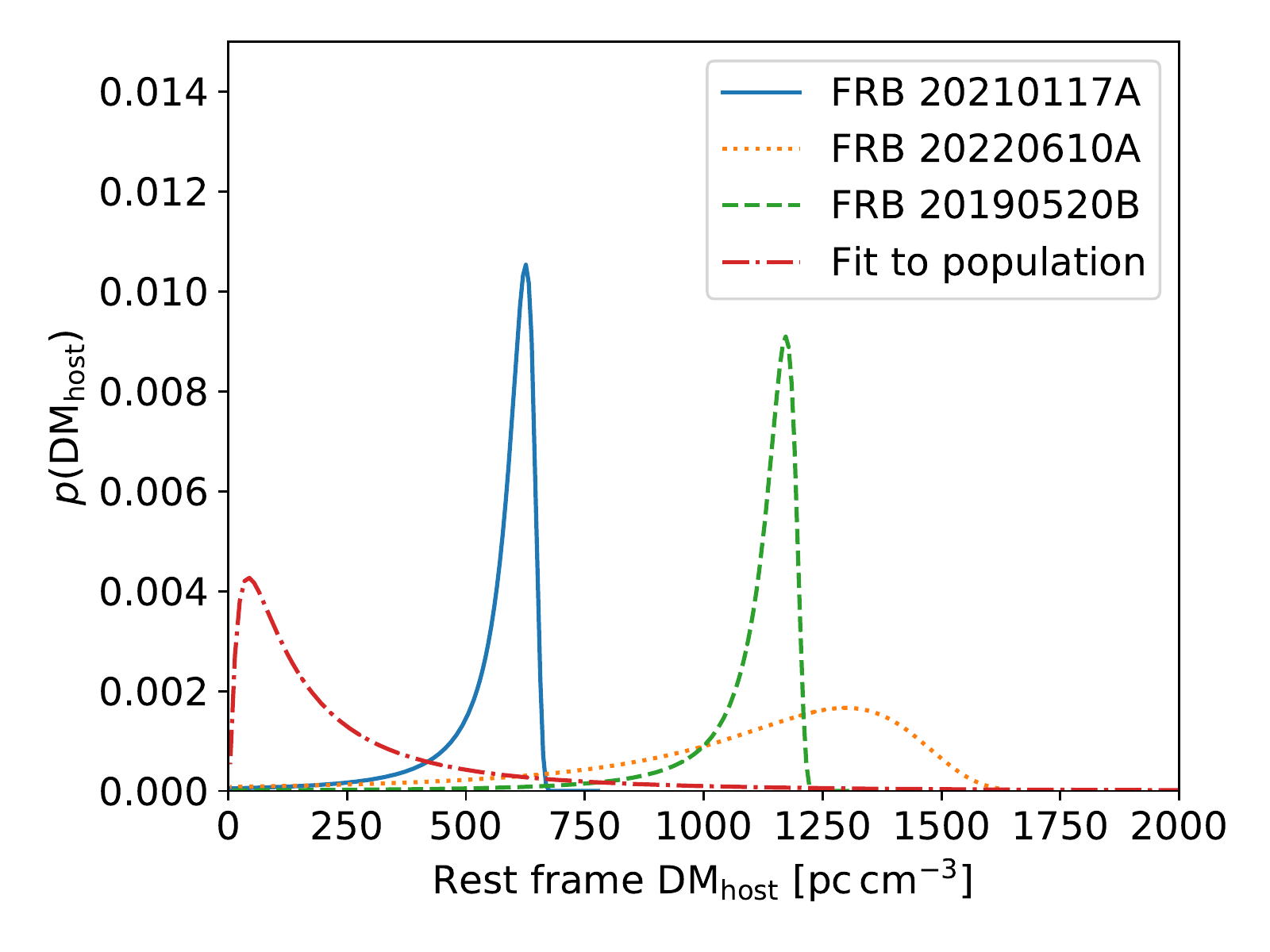} 
\caption{Probability density functions P(DM$_{\rm host}$) for the host galaxy DM contribution, scaled by $1+z$ to the host galaxy's rest frame. Shown are values for three localized FRBs --- blue solid: FRB~20210117; green dashed: FRB~20190520B \citep{Niu+22}; orange dotted: FRB~20220610A \citep{Ryder+22} --- and the log-normal fit to the FRB population based on ASKAP and Parkes (Murriyang) data \citep{James+22}.}
\label{fig:pdmhost}
\end{figure}

\begin{figure}
\includegraphics[width=0.5\textwidth]{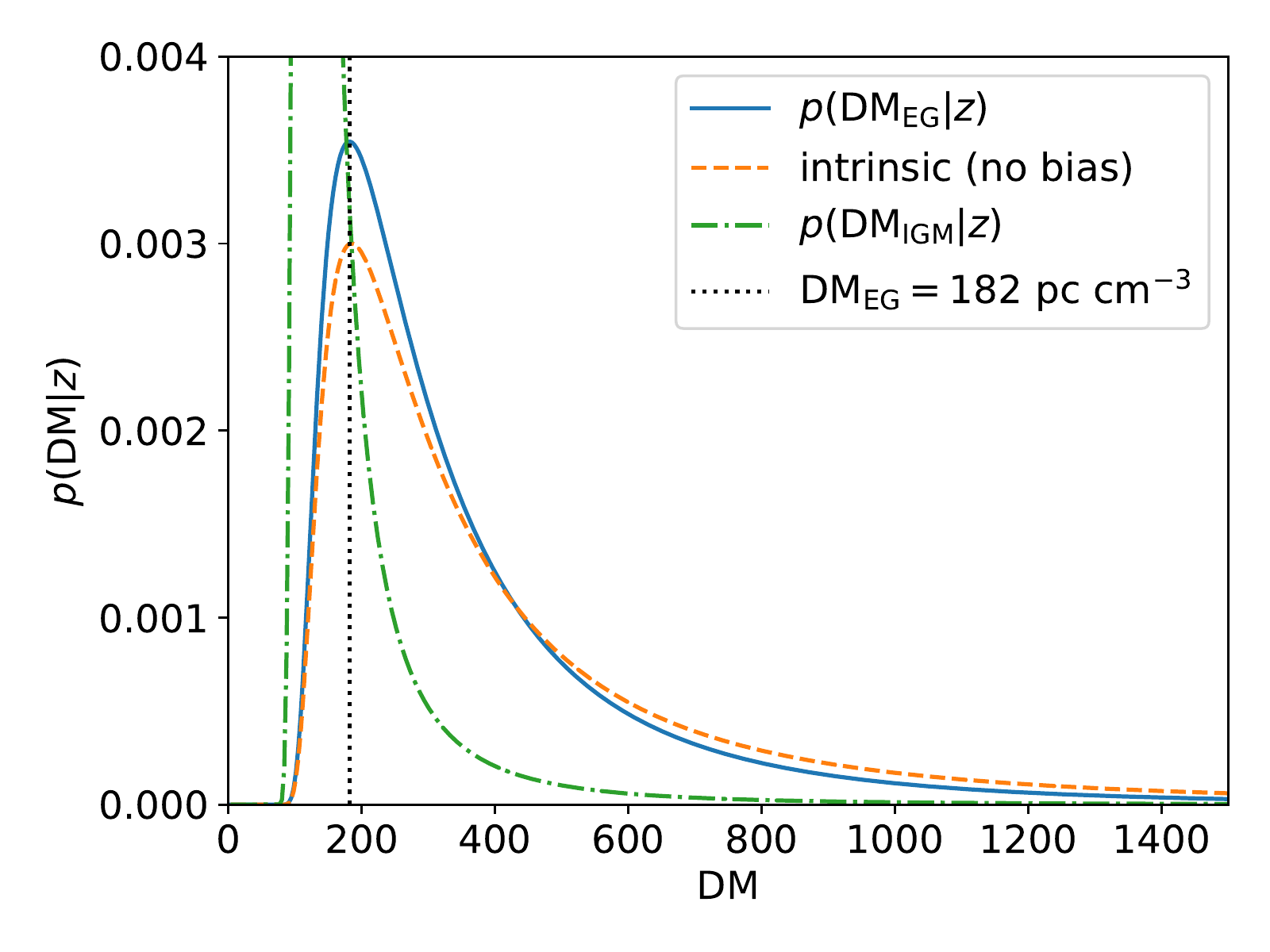} 
\caption{ Probability density function (pdf) P(DM$_{\rm EG}\mid$z) for the extragalactic contribution to the DM given the redshift of the FRB host galaxy. The ASKAP ICS survey yields a blue curve after accounting for various survey biases (peak shown by vertical dashed line). The green curve represents the pdf of the DM due to the IGM only (i.e.\ without the host galaxy), and the orange curve the extragalactic DM free of instrumental biases.}
\label{fig:dmz_host}
\end{figure}

\begin{figure}
\includegraphics[width=0.45\textwidth]{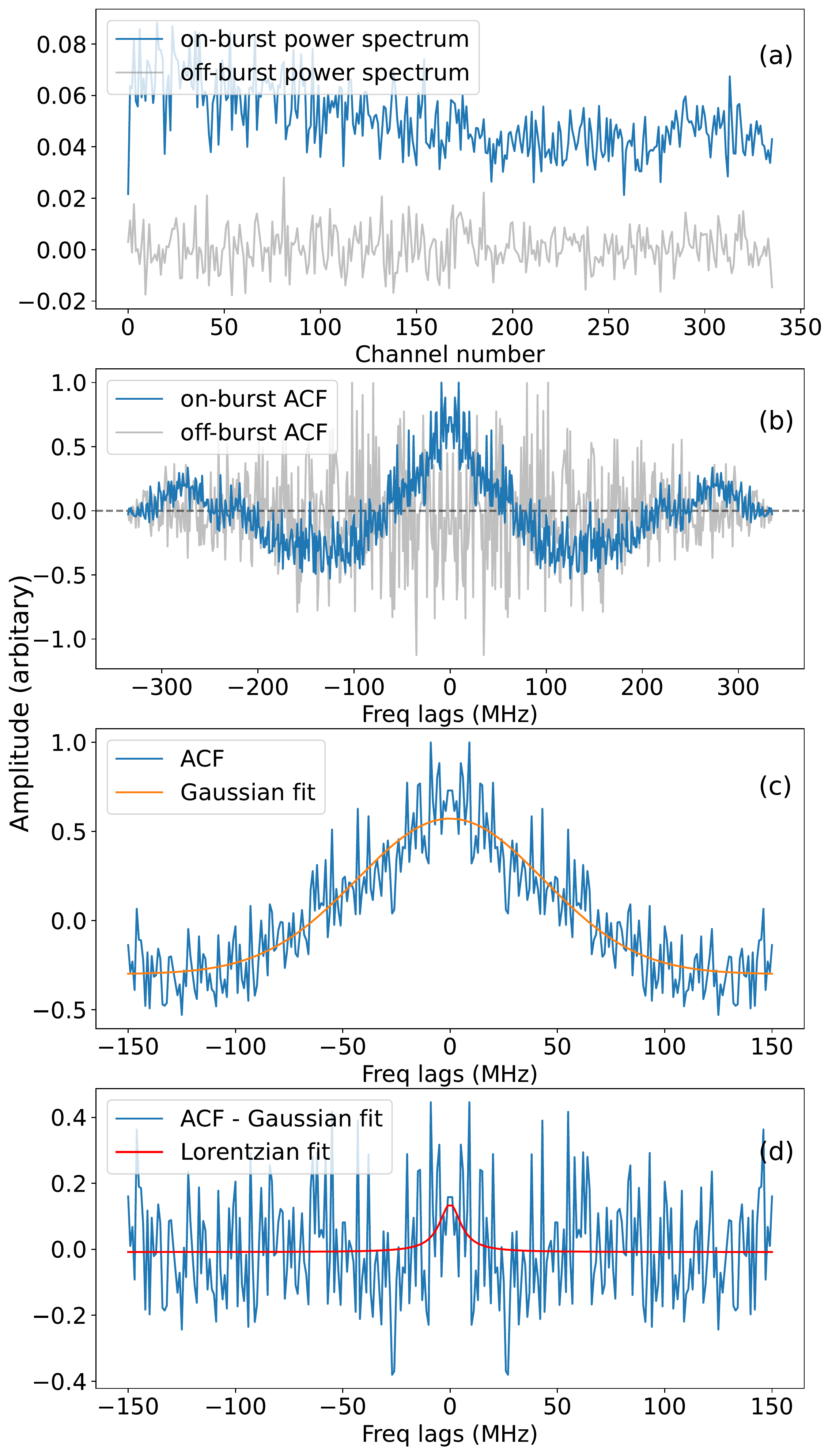} 
\caption{Autocorrelation function (ACF) analysis of the time-averaged spectrum of the FRB (resolution = 1 MHz). Panel (a) shows the burst on and off-peak power spectrum. On and Off-peak ACF function is shown in the panel (b). The noise spike with zero lag has been removed.
A zoomed-in peak of the ACF fitted with a one-component Gaussian function is shown in panel (c). Panel (d) shows the Lorentzian fit to the residual.  }
\label{fig:ACF}
\end{figure}

\section{High-time resolution studies}
Using the CRAFT voltage data, we performed a high-time resolution analysis of the FRB.
The data were beam-formed (coherently summed) at the position of the burst using the delay, bandpass, and phase solutions derived from the calibrator source PKS 0407$-$658.
The 336 1-MHz bandwidth ASKAP channels containing FRB signal were then coherently de-dispersed at the FRB's structure-maximized DM and passed through a synthesis filter to reconstruct a single 336-MHz channel with $\sim$3\,ns time resolution. \citet{Cho+20} provides a detailed description of the high time resolution construction process.  

Next, we characterise the spectral modulation in the burst, which could be intrinsic to the burst emission, or be caused by propagation effects. The autocorrelation function (ACF) of the main component of the burst spectra $S(\Delta \nu$) with a frequency resolution of 1\,MHz is calculated as follows: 
\begin{equation}
    A(\delta \nu) = \frac{1}{N} \sum_{\nu} \Delta S(\nu) \Delta S(\nu + \delta \nu),
\end{equation}
where $\Delta S(\nu) = S(\nu) - \bar{S}$, $\bar{S}$ is the mean spectral power and $N$ is the number of frequency bins \citep{1966AJ.....71R.869S}. The ACF was then normalised by its maximum and fitted with a one-component Gaussian function from the \soft{lmfit} python package. The central peak Full-Width at Half Maximum (FWHM) is $103 \pm 4$\,MHz which is the characteristic frequency scale seen on the spectrum of burst's main component. The fitted Gaussian function is then subtracted from the ACF, and the residuals are fitted with a Lorentzian function of the following form: 

\begin{equation}
    f(\delta \nu) = C\left(1+\frac{\delta \nu^{2}}{\delta \nu_{\rm d}^{2}}\right)^{-1},
\end{equation}
where $C$ is a constant and $\delta \nu_{\rm d}$ is the scintillation bandwidth (see Fig\,\ref{fig:ACF}). We estimate $\delta {\nu}_{\rm d} \sim 6$\,MHz, which is consistent with the expectations for diffractive scintillation (DISS) from the Milky Way along the burst line of sight using the NE2001 model ($\sim 3$\,MHz at 1\,GHz).

We fit the frequency-averaged pulse profile with scatter broadened Gaussian pulse models using nested sampling presented in \citet{Qiu20} and \citet{Cho+20}. This allows the fitting of multiple pulse components within the spectrum as demonstrated in \citet{Day2020}. 

We model the burst using a three-component scattered pulse with a precursor component (see Figure\,\ref{fig:scattering}). Fitting of the averaged pulse profile gives a scattering time of $\tau~=~0.33\pm 0.02$\,ms at centre frequency of 1271\,MHz, assuming $\tau \propto \nu^{-4}$. We note that the scattering fit was performed on the dynamic spectra dedispersed at the structure-maximized DM. We also estimate scattering time as a function of different DM trials between $728.6-729.4$\,pc\,cm$^{-3}$ and find a gradient of $-76$\,$\upmu$s per pc\,cm$^{-3}$. We note that the scattering timescale is not consistent with the MW (0.06\,$\mu$s at 1\,GHz) estimate from NE2001 model. Peak 1 and peak 2 have widths of $0.14 \pm 0.01$\,ms and $0.17 \pm 0.02$\,ms, respectively. The two peaks of the main pulse are separated by 0.60\,ms. 
The precursor emission peak occurs $\sim$1.5\,ms before the main peak with a pulse width of $0.53 \pm 0.03$\,ms. 
\begin{figure}
\includegraphics[width=0.5\textwidth]{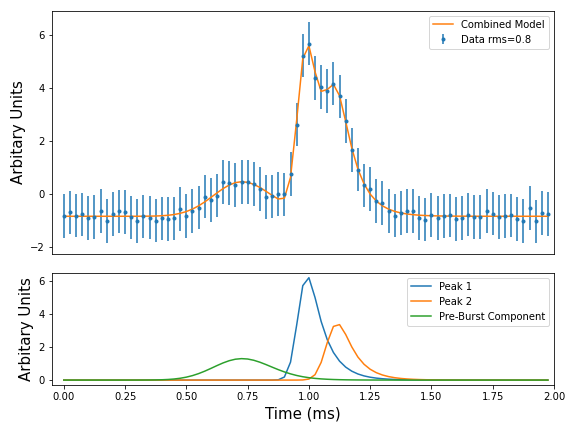} 
\caption{Pulse morphology model plot over 125 $\mu\rm{s}$ time series. The best-fit model comprises a scattered three component pulse profile .}
\label{fig:scattering}
\end{figure}

\subsection{Polarimetry}
The FRB data were polarisation calibrated using an observation of the Vela pulsar (PSR J0835$-$4510), which was observed 3.4\,hr after the detection of the FRB. This ASKAP observation was compared to a Parkes radio telescope observation of the Vela pulsar with an accurate polarisation calibration to determine ASKAP's instrumental leakage parameters (differential gain and phase between the two linearly polarized receptors), which were then applied to the burst data set. See \citet{Day+21} for additional details. 

We used the \texttt{RMFIT} program in \texttt{PSRCHIVE} to calculate the rotation measure (RM) of FRB\,20210117A and find the burst RM to be $43\pm 0.6$\,rad\,m$^{-2}$. The frequency-integrated burst profiles (corrected for Faraday rotation) and the dynamic spectra are presented for all four Stokes parameters, in Figures \ref{fig:profiles} and \ref{fig:dynspecs} respectively.
We see a hint of downward drifting structure in the dynamic spectrum, which has now been established as a distinguishing feature of repeating FRBs \citep{Pleunis+21}. Next, we use the method described in \S2.4.1 of \cite{Day2020} to calculate the polarisation position angle (PA) $\Psi$ and the associated uncertainty $\sigma_\Psi$, which was estimated using the Faraday-corrected Stokes profiles $I$, $Q$, and $U$. The uncertainties $\sigma_I$, $\sigma_Q$, and $\sigma_U$ were estimated by taking the standard deviation of the off-burst Stokes $I$, $Q$, and $U$ data. 
The PA and associated error is shown in the top panel of Fig\,\ref{fig:profiles}. 

We also calculate polarisation fractions for FRB\,20210117A time window of 3.6\,ms using the calibrated Stokes parameters. The total de-bias linear polarisation and total polarisation is given by:
\begin{equation}
  L_{{\textrm{de-bias}}} =
    \begin{cases}
      \sigma_I \sqrt{\left(\frac{L_{{\rm meas}}}{\sigma_I}\right)^2 - 1} & \text{if $\frac{L_{{\rm meas}}}{\sigma_I} > 1.57$} \\
      0 & \text{otherwise} . 
    \end{cases} \\
\end{equation}
\begin{equation}
P  = \sqrt{L_{\textrm{de-bias}}^2 + V^2} \label{eq:P} 
\end{equation}
Here, L$_{\rm meas} = \sqrt{Q^2+U^2}$. 
We obtain $L_{\textrm{de-bias}}/I = 0.90$, $V/I = -0.03$ and $P/I = 0.90$.

\begin{figure}
    \centering    
    \includegraphics{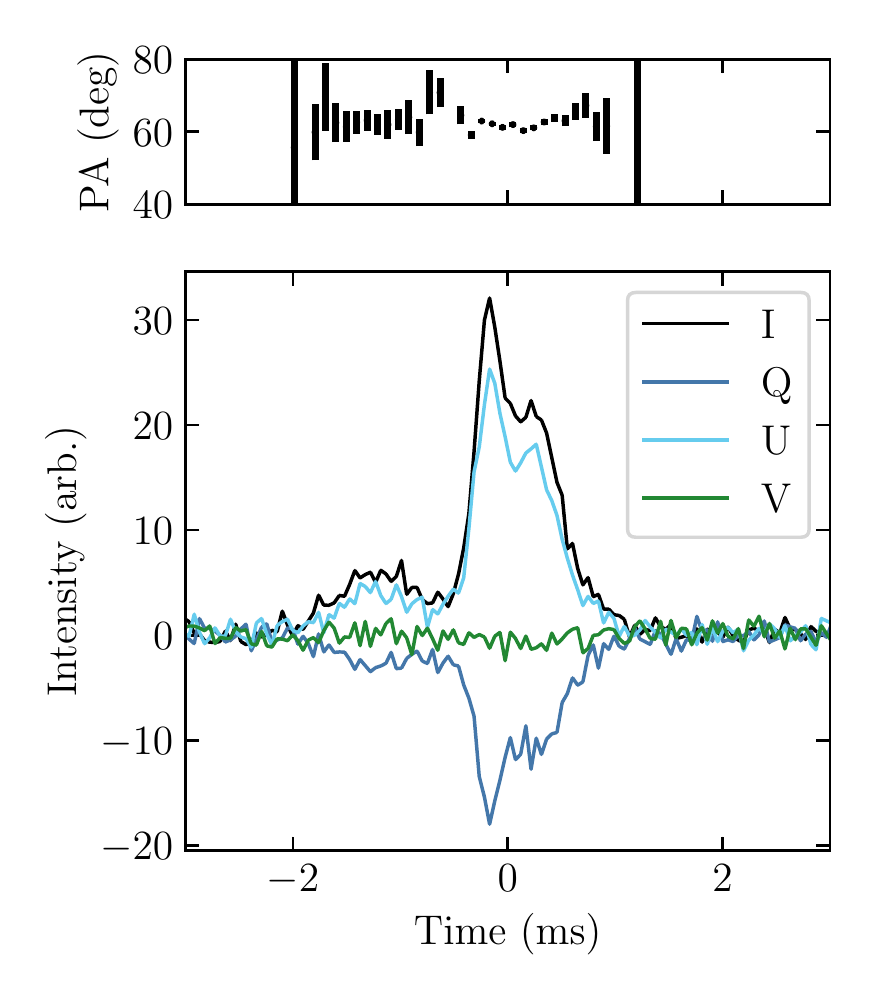}
    \caption{Faraday-corrected profiles of FRB20210117A. Top: polarisation position angle at a time resolution of 96\,$\mu$s. Bottom: Stokes profiles at a time resolution of 48\,$\mu$s.}
    \label{fig:profiles}
\end{figure}
\begin{figure}
    \centering
    \includegraphics{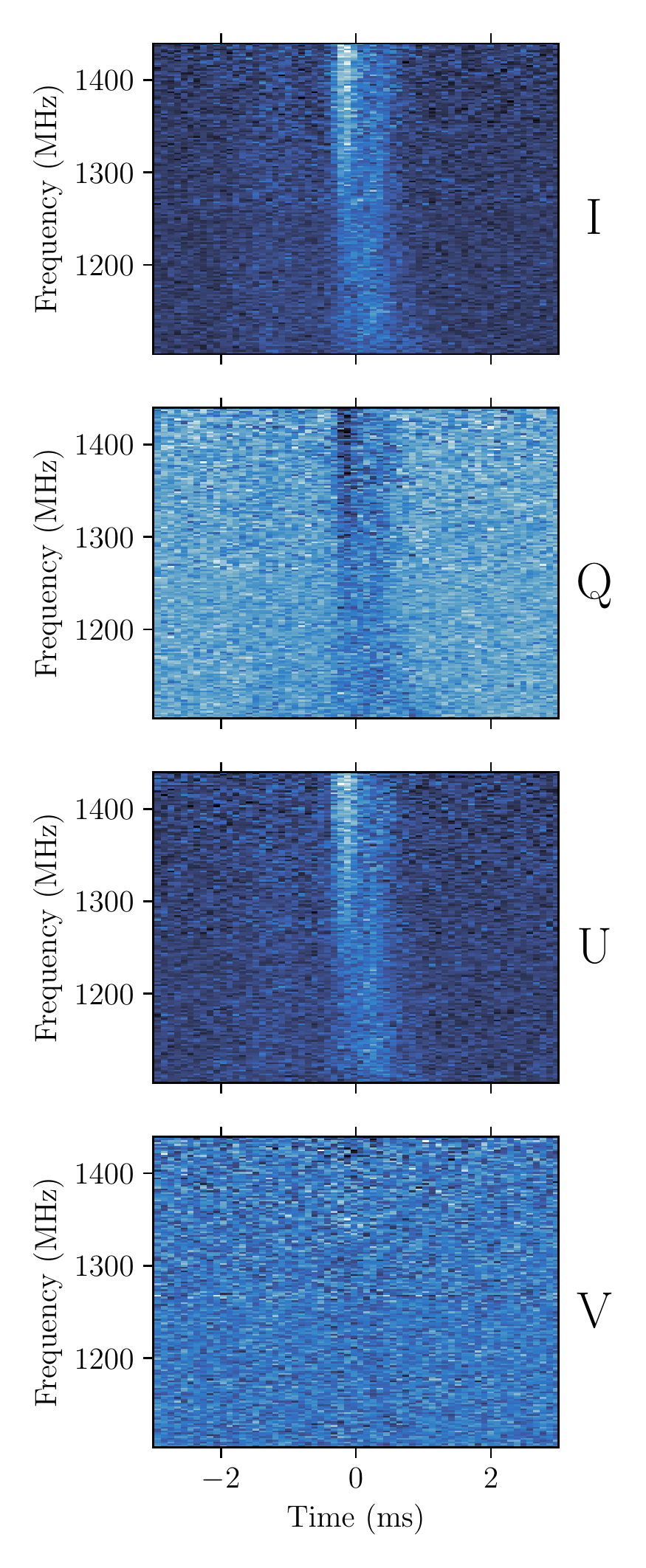}
    \caption{Faraday-corrected Stokes dynamic spectra of FRB20210117A at a time resolution of 96\,$\mathrm{\mu s}$ and frequency resolution of 2\,MHz.}
    \label{fig:dynspecs}
\end{figure}

\section{Follow-up radio observations}
\subsection{Search for a radio persistent source}
We observed the FRB field with the Karl G. Jansky Very Large Array (VLA) under the project code VLA/20B-103 on 2021 February 20. The source was observed for 52\,minutes in $4-8$\,GHz frequency band centered at 6\,GHz. We also conducted second epoch of follow-up observation with the   
Australia Compact Array Telescope (ATCA) on 2021 September 10 for $\sim 3$\,hrs centered at 5.5\,GHz and 7.5\,GHz. We found no radio emission from anywhere in the host galaxy and no compact persistent radio source at the position of the burst. Our $3\sigma$ luminosity limits are $1.5\times 10^{21}$\,W\,Hz$^{-1}$ at 6\,GHz and $4.0\times 10^{21}$\,W\,Hz$^{-1}$ at 6.5\,GHz for both epochs  respectively. These limits are lower than the luminosity of the FRB\,20121102A PRS (see Table\,\ref{tab:case}), indicating that FRB\,20210117A may be an older source or in a less dense environment. 

\subsection{Search for repeating bursts}
We conducted follow-up observations of FRB\,20210117A using the ultra-wideband low (UWL) receiver at the 64-m Parkes radio telescope (also known as \textit{Murriyang}). The observations were centred at 2368 MHz, with the bandwidth spanning $0.7-4$\,GHz. The FRB source was observed for a total of 9.2 hours during January and October 2021. We searched the Parkes data for repeat bursts and single pulses using the \soft{Heimdall} \citep{Barsdell:2012PhDT} and \soft{Fetch} \citep{Agarwal:2020} software packages for a DM range of 100--1100\DMunits, utilizing a tiered sub-band strategy as described in \citet{Kumar:2021}. No significant single-pulse candidates of astrophysical origin were identified in these observations above an S/N of 8. We can constrain the detectable fluence of the repeat bursts to be $\lesssim$ 0.15 Jy\,ms in these UWL observations assuming a broadband pulse (3.3 GHz bandwidth) pulses with a nominal width of 1\,ms. If the repeat bursts are narrowband (64 MHz bandwidth), in that case, our search pipeline was sensitive up to $\sim$1 Jy\,ms. Furthermore, the source was self-followed up with ASKAP between September 2021 to January 2022 for a duration of 125.53 hours, with the band centre frequency ranging from 920.5--1632.5\,MHz. No significant candidates for repeat bursts were found in these ASKAP observations exceeding a threshold S/N of 10. The ASKAP detection system in the incoherent sum mode is sensitive to a fluence of 3.7 Jy\,ms for a nominal pulse width of 1 ms using the entire array of 36 antennas. Although, for most of the follow-up observations, smaller sub-arrays were used consisting of 23--26 antennas. Assuming a Poissonian rate distribution, we set a 95\% upper limit on the burst repetition rate to be $\sim2.4\times10^{-2}$\,hr$^{-1}$ for ASKAP observations. We note here that in some repeating FRB sources, the burst rate has been found to show significant variations with time as well as frequency \citep{Cruces:2021, Xu:2022, Dai+22}, and so this upper limit is just a rough estimate for repetition.

\section{Discussion}
\subsection{Comparison with FRB host population}
\begin{figure}
\includegraphics[scale=0.4]{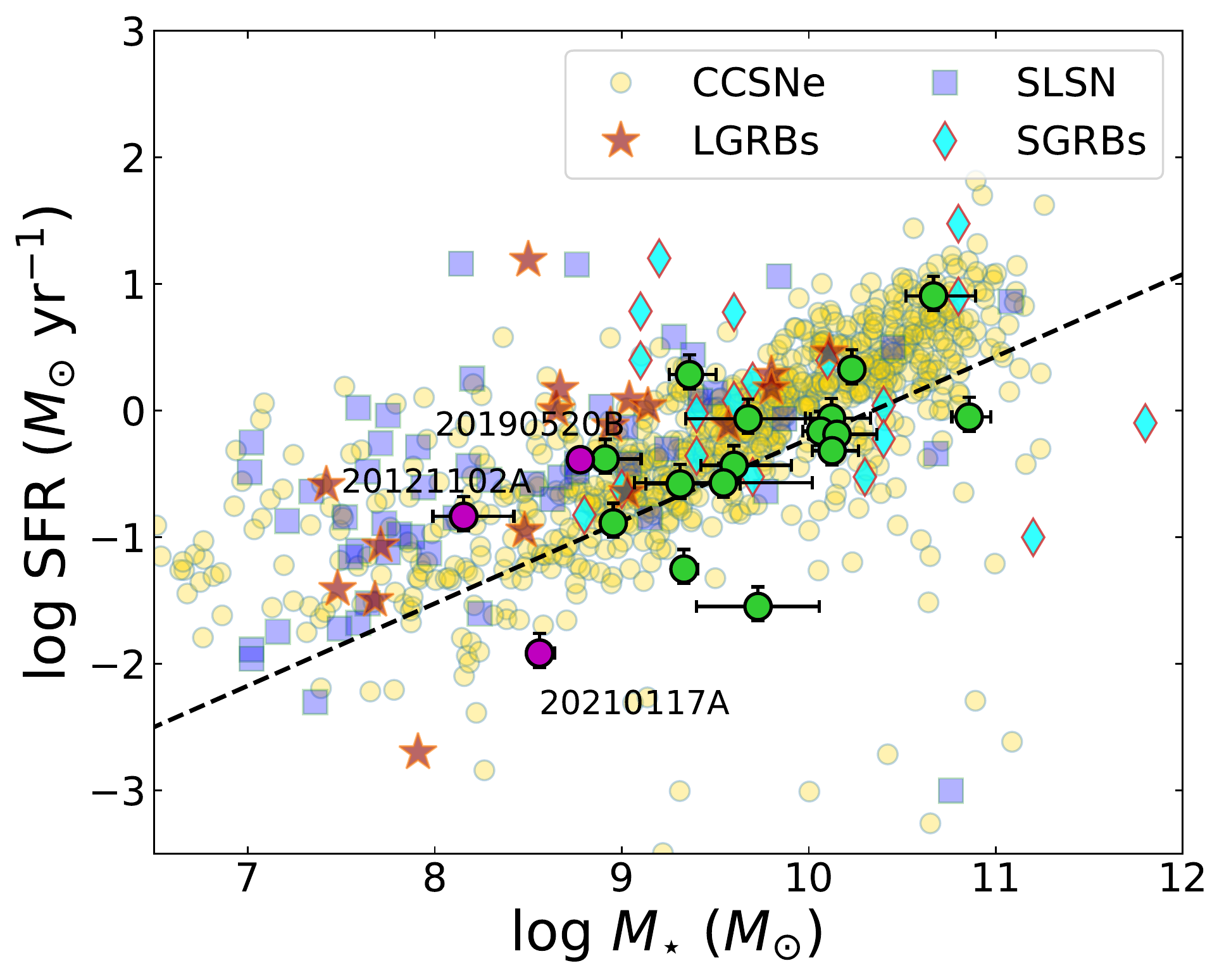} 
\caption{Stellar mass and SFR distribution for FRB hosts compared hosts of other transients. FRB dwarf host galaxies are shown in magneta, while other FRB hosts are shown in green. The hosts of CCSNe (yellow circle), SLSNe (blue squares), SGRBs (cyan diamonds), and LGRBs (red stars) are overplotted. Dashed line separates the star-forming and quiescent galaxies. The data for transients is taken from \citet{Taggart+21}, \citet{Schulze+20} and \citet{Nugent+22}.}
\label{fig:compareFRB}
\end{figure}


FRB\,20210117A is the \emph{only} published burst discovered in a dwarf galaxy where the repeating nature has not yet been established. We compare this source with the published sample of FRB hosts, particularly with FRB\,20121102A and FRB\,20190520B, which are known to originate in dwarf galaxies. Despite the large excess host DM, the local environment of FRB\,20210117A differs from those of FRB\,20121102A and FRB\,20190520B due to the lack of a PRS and a low rotation measure which is suggestive of low magnetic fields or an older system (see Table\,{\ref{tab:case}}). 
\begin{table*}
    \small
	\caption{Observed properties of FRB\,20210117A along with that of two active repeating FRBs localized to dwarf galaxies. The burst rate for 20121102A is the peak rate at 1.25\,GHz above a fluence of 0.0015 Jy\,ms and R = $_{\rm 1.2\,GHz} (>9.3$\,mJy\,ms) for 20190520B. We quote the radio luminosities for the PRS at 5\,GHz. The properties of FRB\,20121102A are taken from \citet{Li+21,MichilliRM,Tendulkar17,VLAlocalisation,Marcote17, Hessels19ApJL} and those for FRB\,20190520B are taken from \citet{Niu+22,Dai+22,Anna-Thomas+22}  }
    \begin{tabular}{ccccccccc}
\hline
FRB & $z$ & Repeat rate & RM & PRS luminosity & Host DM & Scattering ($\tau_{1\,\rm GHz})$ & Pulse morphology\\
&  & (hr$^{-1}$)& (rad\,m$^{-2}$) & (W\,Hz$^{-1}$) & (pc\,cm$^{-3}$) & (ms) & \\
\hline
20121102A & 0.192 & 122 & $10^{5}$ & $1.4 \times 10^{22}$ & $\leq324$ & 0.02 & repeater-like \\
20190520B & 0.241 &  $4.5^{1.9}_{-1.5}$  & $10^{3}-10^{4}$ & $2\times 10^{22}$ & $903^{+72}_{-111}$  & 24.4 &repeater-like \\
20210117A & 0.214 & $<2.4\times 10^{-2}$ & $43$ & $<5.3 \times 10^{21}$ & $\sim$460 & 0.86 &repeater-like \\
\hline
\end{tabular}
\label{tab:case}
\end{table*}

We note that FRB\,20210117A has the second highest excess DM in the sample of ASKAP-localized bursts after FRB\,20220610A. Simha et al. (in prep.) also studied the matter density distribution along the FRB sightline and discovered no foreground galaxies or haloes to explain the excess DM. Their study leveraged the spectroscopic redshifts of field galaxies from the FLIMFLAM survey (see \citet{Lee+22}) to model the foreground gas distribution from intervening galactic halos and also searched for possible galaxy groups whose inter-group medium might contribute to the DM. The resulting empirical model of foreground plasma indicates a very small contribution ($<10$\,pc\,cm$^{-3}$) to the DM and thus indicates a high host galaxy or progenitor environment value.

We also search for any possible relationship between the scattering and excess DM for a sample of FRBs including
FRB\,20220610A ($\tau_{1\,\rm GHz}$~=~0.89\,ms) \citep{Ryder+22} and those presented in Table\,{\ref{tab:case}}. 
Except for FRB\,20121102A, the scattering timescales for FRBs in this sample exceed the expectations from the ISM in our galaxy, implying that scattering originates far beyond the Milky Way, possibly in the FRB host galaxy. We also do not find any correlation between scattering timescale and excess host DM estimates. 

In Figure \,\ref{fig:compareFRB}, we compare the stellar mass and SFR of FRB\,20210117A's host with the FRB host population and discover that the host has a very low SFR compared to the population. It is evident that 1) there is little ongoing star formation and no bursts of star formation in the past and, 2) the host DM constraints from H$\alpha$ measurements rule out excess DM contribution from the host, implying that the majority of the observed excess DM must come from the immediate surroundings of the FRB source. In the same figure, we also compare the stellar mass and SFRs with the hosts of other transients such as core collapse supernovae (CCSNe) \citep{Schulze+20}, superluminous supernovae (SLSNe) \citep{Taggart+21}, and long and short GRBs \citep{Taggart+21,Nugent+22}. Unlike dwarf hosts of repeating FRBs, the properties of FRB\,20210117A hosts do not match those of SLSNe and LGRB hosts. Furthermore, unlike hosts of other ASKAP-localized FRBs, host of FRB\,20210117A does not share the same space as the majority of SGRB hosts. However, it is broadly consistent with CCSNe hosts. 
\subsection{Potential progenitor channel} \label{subsec:progenitor}

\begin{figure*}
\includegraphics[width=1\textwidth]{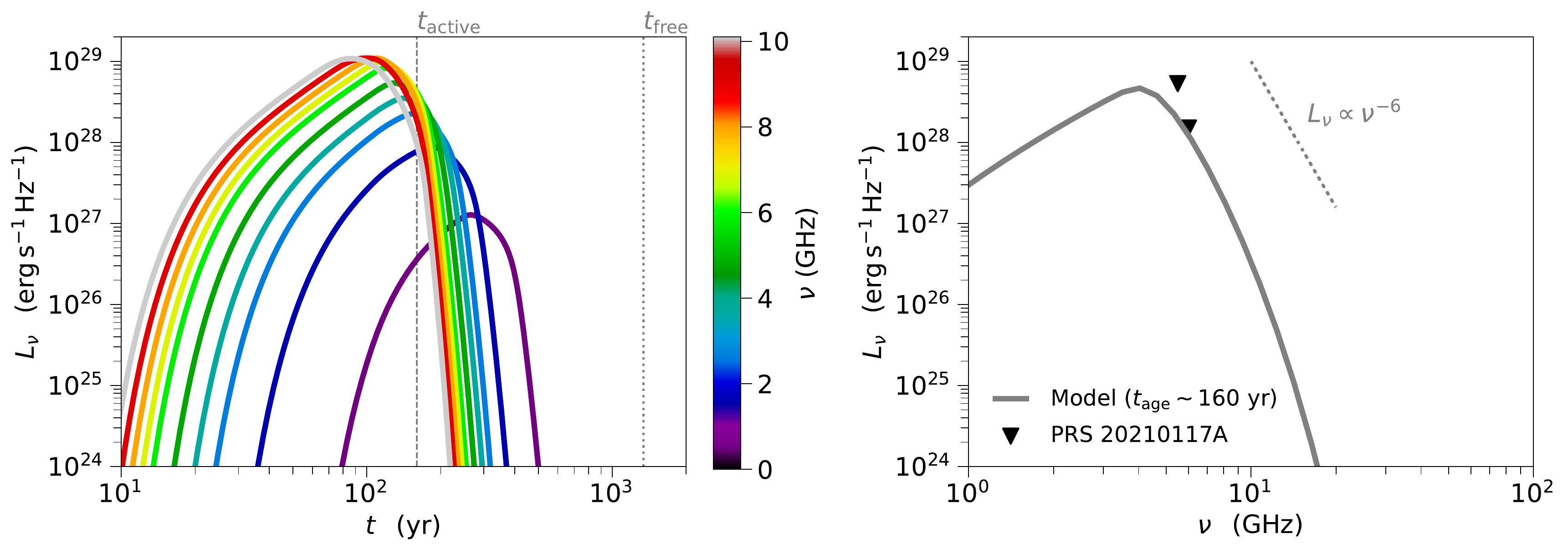}
\caption{Radio synchrotron emission from an accretion-powered hypernebula surrounding the FRB source. Left panel: Light curves of the expanding hypernebula in different bands (color-coded). Vertical grey dashed and dotted lines denote the active duration of the central accreting engine $t_{\rm active}$, and the free-expansion timescale of the hypernebula (Eq.~\ref{eq:t_free}), respectively (see Sec.~\ref{subsec:progenitor} for more details on the system's parameters). Right panel: Spectral energy distribution at an epoch $t_{\rm age}=160$\,yr. Black downward-facing triangles are upper-limits of the persistent radio source (PRS) associated with FRB\,20210117A.}
\label{fig:PRSmodel}
\end{figure*}
The large DM$_{\rm host}$ of FRB\,20210117A strongly hints at the existence of a compact nebula surrounding the FRB engine. Such sources could be powered, for instance, by young pulsars in supernova remnants \citep{Piro+16,Connor+16}, or interactions of strong winds from a young magnetar with the surrounding medium to form a pulsar wind nebula \citep{Dai+17, Margalit+18}. The recently proposed scenario
whereby the FRB source is embedded within the powerful baryon-rich outflows from a hyper-accreting black hole \citep[`hypernebula';][]{Sridhar&Metzger22, Sridhar+23b} could explain various properties of FRB\,20210117A, including its large DM$_{\rm host}$. We further investigate this model in light of our observations. 

The derived isotropic-equivalent luminosity of the burst seen from FRB\,20210117A is $L_{\rm FRB}=1.36^{+1.06}_{-0.34}\times10^{43}$\,erg\,s$^{-1}$ 
This requires a minimum accretion rate of $\dot{m}=\dot{M}/\dot{M}_{\rm Edd}\gtrsim10^6$ for the FRB to be accretion-jet powered \citep{Sridhar+21b}, where $\dot{M}_{\rm Edd}$ is the Eddington mass transfer rate for an accreting $10M_{\odot}$ black hole. Such accretion-jet powered scenario could give rise to repeating, and potentially, even periodically active FRBs, where the periodicity may be asssociated with the Lens-Thirring precession timescale of the accretion disk/jet passing along the observers' line of sight. The apparent non-repetition from FRB\,20210117A, in this scenario, could imply a small activity duty cycle \citep{Sridhar+21b, Katz_21}, 
\begin{equation} \label{eq:duty_cycle}
\zeta \approx\left(\frac{4f_{\rm b}}{\pi^2\theta_0^2}\right)^{1/2} \approx \frac{5}{\dot{m}\theta_0} = 5\times10^{-5}\left(\frac{\dot{m}}{10^6}\right)^{-1}\left(\frac{\theta_0}{0.1}\right)^{-1},
\end{equation}
where $f_{\rm b}=2\pi(1-\cos{(\Delta\theta)})/4\pi\sim0.01$ is the FRB beaming factor, $\theta_0$ is the angle of the axis of jet precession, and $\Delta\theta$ is the opening angle of the jet.

The quasi-spherical disk winds, as they expand, drive a forward shock into the circumstellar medium with a typical density of say, $n\approx10\,{\rm cm}^{-3}$. On the other hand, the faster jet interacts with the slower disk winds via a termination shock. Following \cite{Sridhar&Metzger22}, we calculate the observable properties of the hypernebula due to these interactions for the following physical parameters: velocity of the slower disk wind $v_{\rm w}=0.01\,c$, velocity of the fast wind/jet $v_{\rm j}=0.1\,c$, jet magnetization parameter (ratio of the magnetic energy density to the plasma rest mass energy density) $\sigma_{\rm j}=0.1$, ratio of the wind luminosity to the jet luminosity $\eta=0.1$, fraction of the shock power that goes into heating the electrons $\varepsilon_{\rm e}=0.5$, mass of the accreting black hole $M_\bullet=10 M_\odot$, and mass of the companion accretor star $M_\star=30 M_\odot$. The free expansion timescale of the outflowing winds (before they start to decelerate) is,
\begin{equation} \label{eq:t_free}
    t_{\rm free} \approx 1.3\times10^3\,{\rm yr}\left(\frac{L_{\rm w,42}}{n_1}\right)^{1/2}\left(\frac{v_{\rm w}}{0.01\,c}\right)^{-2.5}.
\end{equation}
Above, we adopt the short-hand notation, $Y_{x} \equiv Y/10^x$ for quantities in cgs units.

During the free-expansion phase, the ionized wind shell contributes to a dispersion measure through it given by \citep{Sridhar&Metzger22},
\begin{equation}
    {\rm DM}_{\rm sh} \simeq 470\,{\rm pc\,cm}^{-3}\left(\frac{\dot{m}}{10^6}\right)\left(\frac{v_{\rm w}}{0.01\,c}\right)^{-2}\left(\frac{t_{\rm age}}{160\,{\rm yr}}\right)^{-1}.
\end{equation}
The model prediction of ${\rm DM}_{\rm sh}\simeq470$\,pc\,cm$^{-3}$ is consistent with the observed ${\rm DM}_{\rm host}\sim460$\,pc\,cm$^{-3}$ 
for our chosen set of parameters. For the same set of parameters, the left panel of Fig.~\ref{fig:PRSmodel} shows the model light curves in different bands (0.1--10\,GHz). Also indicated there are the active duration of the engine $t_{\rm active} \approx M_\star/\dot{M}\simeq150$\,yr, and $t_{\rm free}$. Shown in the right panel of Fig.~\ref{fig:PRSmodel} is the model spectrum of the radio synchrotron emission from the shock-heated electrons, calculated when the age of the hypernebula is $t_{\rm age}\sim$160\,yr (expansion timescale)---corresponding to the time when the model ${\rm DM}_{\rm sh}$ tentatively matches the observed ${\rm DM}_{\rm host}$. We note here that the model spectrum calculated at this time of expansion is also in agreement with the upper limits on the observed persistent radio emission from FRB\,20210117A.

The absolute maximum rotation measure through the nebula at time $t_{\rm age}\sim160$\,yr is $|{\rm RM}|_{\rm max}\simeq2\times10^7$\,rad\,m$^{-2}$ \citep[Eq.~50 of][]{Sridhar&Metzger22}. The observed RM of $\sim$40\,rad\,m$^{-2}$ is thus consistent within this model and could mean that the FRB was observed during a phase of RM sign reversal, as known to be seen from other FRBs too \citep{Anna-Thomas+22, Mckinven+22, Dai+22}. Such RM swings could be due to fluctuating orientations of local magnetic field lines in the turbulent eddies downstream of the termination shock, as can be expected from accreting BH outflows (e.g., Eq.~51 of \citealt{Sridhar&Metzger22}; see also \citealt{Yang+22}). Future long-term, short-cadence observations will reveal the trend of $|{\rm RM}|_{\rm max}$ and allow us to constrain the model parameters better to consistently explain the observed RM(t) as well as the spectra.

\section{Summary}
We have presented the discovery and sub-arcsecond localization of an apparently one-off FRB\,20210117A which originates in a dwarf galaxy at a $z=0.214$. The dwarf host galaxy has a little ongoing starformation as compared to the known dwarf host of repeating FRBs. FRB\,20210117A is among the sample of FRBs with an excess host DM contribution (DM$_{\rm host} \sim 460$\,pc\,cm$^{-3}$), where the excess DM is more likely to come from the burst's local environment. The burst is highly (90\%) linearly polarised, has a low rotation measure (RM$=43$\,rad\,m$^{-2}$) and a flat polarisation position angle. A high time resolution analysis of FRB\,20210117A and its dynamic spectrum reveals that the burst has three components and a hint of frequency drifting. While none of these characteristics are inconsistent with a non-repeating origin, flat polarization position angles and frequency drifting in particular are more commonly found in repeating sources; however, subsequent observations have not detected any repeat bursts. Moreover, we find no radio emission (either a PRS or from star formation) in our follow-up observations. Thus, the local environment of FRB\,20210117A is very different from repeating FRBs 20121102A and 20190520B with a dwarf host galaxy. Finally, we find that accretion-jet powered hypernebula model for FRB\,20210117A matches with our observations. 

We encourage follow-up observations to search for repeating pulses. The discovery of a repeating burst from FRB\,20210117A would contradict the observed correlation between FRBs originating in dwarf galaxies and their association with a PRS.

\vskip 5mm
SB would like to thank Elizabeth A. K. Adams, Reynier Peletier, Jason Hessels and the Astroflash group for useful discussions. 
SB is supported by a Dutch Research Council (NWO) Veni Fellowship (VI.Veni.212.058). J.X.P. as a member of the Fast and Fortunate for FRB Follow-up team, acknowledge support from 
NSF grants AST-1911140, AST-1910471 and AST-2206490.
K.W.B., J.P.M, and R.M.S. acknowledge Australian Research Council (ARC) grant DP180100857. RMS acknowledges support through Australian Research Council Future Fellowship FT190100155 and Discovery Project DP220102305. T.E. is supported by NASA through the NASA Hubble Fellowship grant HST-HF2-51504.001-A awarded by the
Space Telescope Science Institute, which is operated by the
Association of Universities for Research in Astronomy, Inc.,
for NASA, under contract NAS5-26555. N.S. acknowledges support from NASA (grant number 80NSSC22K0332), NASA FINESST (grant number 80NSSC22K1597), and Columbia University Dean's fellowship. C.W.J.\ and M.G.\ acknowledge support by the Australian Government through the Australian Research Council's Discovery Projects funding scheme (project DP210102103).
W.F. and A.C.G. acknowledge support by the National Science Foundation under CAREER grant No. AST-2047919, and by the David and Lucile Packard Foundation.

Based on observations collected at the European Southern Observatory under ESO programme 0105.A-0687. The Australian Square Kilometre Array Pathfinder is a part of the Australia Telescope National Facility which is managed by CSIRO. Operation of ASKAP is funded by the Australian Government with support from the National Collaborative Research Infrastructure Strategy. ASKAP uses the resources of the Pawsey Supercomputing Centre. Establishment of ASKAP, the Murchison Radio-astronomy Observatory and the Pawsey Supercomputing Centre are initiatives of the Australian Government, with support from the Government of Western Australia and the Science and Industry Endowment Fund. 
We acknowledge the Wajarri Yamatji as the traditional owners of the Murchison Radio-astronomy Observatory site. 
The Australia Telescope Compact Array is part of the Australia Telescope National Facility which is funded by the Australian Government for operation as a National Facility managed by CSIRO. We acknowledge the Gomeroi people as the traditional owners of the Observatory site. The National Radio Astronomy Observatory is a facility of the National Science Foundation operated under cooperative agreement by Associated Universities, Inc. Some of the data presented herein were obtained at the W. M. Keck Observatory, which is operated as a scientific partnership among the California Institute of Technology, the University of California and the National Aeronautics and Space Administration. The Observatory was made possible by the generous financial support of the W. M. Keck Foundation. W. M. Keck Observatory access was supported by Northwestern University and the Center for Interdisciplinary Exploration and Research in Astrophysics (CIERA). The authors wish to recognize and acknowledge the very significant cultural role and reverence that the summit of Maunakea has always had within the indigenous Hawaiian community.  We are most fortunate to have the opportunity to conduct observations from this mountain. 

\vspace{5mm}
\facilities{ASKAP, VLA, ATCA, VLT (HAWK-I, FORS2), Keck.}


\software{astropy \citep{Astropy:2013, Astropy:2018},  
          numpy \citep{Harris:2020_numpy}, 
          matplotlib \citep{Hunter:2007_matplotlib},
          lmfit \citep{Newville:2016},
          PyMultiNest \citep{Buchner:2014},
          bilby \citep{Ashton:2019},
          fetch \citep{Agarwal:2020},
          PSRCHIVE \citep{psrchive},
          miriad \citep{miriad},
          CASA \citep{CASA}.
          }
                    
\bibliography{references_new.bib}

\end{document}